\documentclass[11pt]{article}
\usepackage{amsthm}
\newtheorem{fact}{Fact}

\usepackage{amsmath, amssymb}
\theoremstyle{plain}
\newtheorem{theorem}{Theorem}[section]
\newtheorem{lemma}[theorem]{Lemma}

\newtheorem{proposition}[theorem]{Proposition}

\theoremstyle{definition}
\newtheorem{definition}[theorem]{Definition}

\theoremstyle{remark}

\newcommand{\beginproofof}[1]{\medskip\noindent{\bf Proof of #1.~}}
\newcommand{\finishproof}{\hspace{0.2ex}\rule{1ex}{1ex}}

\def\squareforqed{\hbox{\rlap{$\sqcap$}$\sqcup$}}
\def\qed{\ifmmode\squareforqed\else{\unskip\nobreak\hfil
\penalty50\hskip1em\null\nobreak\hfil\squareforqed
\parfillskip=0pt\finalhyphendemerits=0\endgraf}\fi}

\newcommand{\comments}[1]{}

\newcommand{\defeq}{\stackrel{\mathrm{def}}{=}}

\newcommand{\sepAuthor}{1in}
\newcommand{\sepAbstract}{1in}
\newcommand{\skipKeywords}{30pt}
\newcommand{\sepTitle}{2ex}
\newcommand{\bool}{\{0,1\}}
\newcommand{\booln}{\{0,1\}^n}

\long\def\symbolfootnote[#1]#2{\begingroup%
\def\thefootnote{\fnsymbol{footnote}}\footnote[#1]{#2}\endgroup}

\providecommand{\MR}{\relax\ifhmode\unskip\space\fi MR }
 \providecommand{\href}[2]{#2}

\long\def\mytitlepage#1#2#3#4{
        \thispagestyle{empty}
    \vspace*{\sepTitle}
        \begin{center}
        {\Large\bf #1}

        \vspace{\sepAuthor}
        #2\\
        \medskip

        \vspace{\sepAbstract}
        {\Large Abstract}
        \end{center}

        \noindent{#3}
        \vskip\skipKeywords

        \noindent{#4}
        \clearpage
        }

\begin{document}

\mytitlepage{The Communication Complexity of the Hamming Distance
Problem\symbolfootnote[0]{ $^\dagger$Supported in part by NSF
grants 0347078 and 0323555. $^*$Supported in part by NSF grants
CCR-0310466 and CCF-0426582.}}{
Wei Huang$^\dagger$ \quad  \quad Yaoyun Shi$^\dagger$ \quad  \quad Shengyu Zhang$^*$ \quad  \quad Yufan Zhu$^\dagger$\\
       {$^\dagger$Department of Electrical Engineering and Computer Science, University of Michigan}\\
       {1301 Beal Avenue, Ann Arbor, MI 48109-2122, USA}\\
       {Email: weihuang$|$shiyy$|$yufanzhu@eecs.umich.edu}\\
       {$^*$Computer Science Department, Princeton University, NJ 08544,
       USA}\\
       {Email: szhang@cs.princeton.edu}
}{We investigate the randomized and quantum communication
complexity of the \textsc{Hamming Distance} problem, which is to
determine if the Hamming distance between two $n$-bit strings is
no less than a threshold $d$. We prove a quantum lower bound of
$\Omega(d)$ qubits in the general interactive model with shared
prior entanglement. We also construct a classical protocol of $O(d
\log d)$ bits in the restricted {\em Simultaneous Message Passing}
model, improving previous protocols of $O(d^2)$ bits
(A.~C.-C.~Yao, {\em Proceedings of the Thirty-Fifth Annual ACM
Symposium on Theory of Computing}, pp. 77-81, 2003), and $O(d\log
n)$ bits (D.~Gavinsky, J.~Kempe, and R.~de~Wolf, quant-ph/0411051,
2004).}{{\bf Keywords:} Computational complexity, Communication
complexity, Hamming distance}

\section{Introduction}
Communication complexity was introduced by Yao \cite{Yao79} and
has been extensively studied afterward not only for its own
intriguing problems, but also for its many applications ranging
from circuit lower bounds to data streaming algorithms. We refer
the reader to the monograph \cite{KN97} for an excellent survey.

We recall some basic concepts below. Let $n$ be an integer and
$X=Y=\{0, 1\}^n$. Let $f: X\times Y\to\{0, 1\}$ be a Boolean
function. Consider the scenario where two parties, Alice and Bob,
who know only $x\in X$ and $y\in Y$, respectively, communicate
interactively with each other to compute $f(x, y)$. The {\em
deterministic communication complexity} of $f$, denoted by $D(f)$,
is defined to be the minimum integer $k$ such that there is a
protocol for computing $f$ using no more than $k$ bits of
communication on any pair of inputs. The {\em randomized
communication complexity} of $f$, denoted by $R^{pub}(f)$, is
similarly defined, with the exception that Alice and Bob can use
publicly announced random bits and that they are required to
compute $f(x,y)$ correctly with probability at least $2/3$. One of
the central themes on the classical communication complexity
studies is to understand how randomness helps in saving the
communication cost. A basic finding of Yao \cite{Yao79} is that
there are functions $f$ such that $R(f)= O(\log D(f))$. One
example is the \textsc{Equality} problem, which simply checks
whether $x=y$.

Later results show that different ways of using randomness result in
quite subtle changes on communication complexity. A basic finding in
this regard, due to Newman \cite{Newman91}, is that public-coin
protocols can save at most $O(\log n)$ bits over protocols in which
Alice and Bob toss private (and independent) coins. The situation
is, however, dramatically different in the {\em Simultaneous Message
Passing} (SMP) model, also introduced by Yao \cite{Yao79}, where
Alice and Bob each send a message to a third person, who then
outputs the outcome of the protocol. Apparently, this is a more
restricted model and for any function, the communication complexity
in this model is at least that in the general interactive
communication model. Denote by $R^{\|}(f)$ and $R^{\|,
\mathrm{pub}}(f)$ the communication complexities in the SMP model
with private and public random coins, respectively. It is
interesting to note that $R^{\|, \mathrm{pub}}(\textsc{Equality}) =
O(1)$ but $R^{\|}(\textsc{Equality}) = \Theta(\sqrt{n})$
\cite{Ambainis96}\cite{NS96}\cite{BK97}.

Yao also initiated the study of quantum communication complexity
\cite{Yao93}, where Alice and Bob are equipped with quantum
computational power and exchange quantum bits. Allowing an error
probability of no more than $1/3$ in the interactive model, the
resulting communication complexity is the {\em quantum
communication complexity} of $f$, denoted by $Q(f)$. If the two
parties are allowed to share prior {\em quantum entanglement}, the
quantum analogy of randomness, the communication complexity is
denoted by $Q^*(f)$. Similarly, the quantum communication
complexities in the SMP model are denoted by $Q^{\|}$ and $Q^{\|,
*}$, depending on whether prior entanglement is shared. The
following relations among the measures are easy to observe.

\begin{equation}\label{eq: relation}
Q^*(f) \leq \begin{matrix} R^{\mathrm{pub}}(f) \\ Q^{\|, *}(f)
\end{matrix} \leq R^{\|, \mathrm{pub}}(f)
\end{equation}

Two very interesting problems in both communication models are the
power of quantumness, \emph{i.e.} determining the biggest gap
between quantum and randomized communication complexities, and the
power of shared entanglement, \emph{i.e.} determining the biggest
gap between quantum communication complexities with and without
shared entanglement. An important result for the first problem by
Buhrman, Cleve, Watrous and de Wolf \cite{BCW01} is
$Q^{\|}(\textsc{Equality}) = O(\log n)$, an exponential saving
compared to the randomized counterpart result
$R^{\|}(\textsc{Equality}) = \Theta(\sqrt{n})$ mentioned above. This
exponential separation is generalized by Yao \cite{Yao03}, showing
that $R^{\|, \mathrm{pub}}(f) = constant$ implies $Q^{\|}(f) =
O(\log n)$. As an application, Yao considered the \textsc{Hamming
Distance} problem defined below. For any $x, y\in\{0, 1\}^n$, the
Hamming weight of $x$, denoted by $|x|$, is the number of $1$'s in
$x$, and the Hamming distance of $x$ and $y$ is $|x\oplus y|$, with
``$\oplus$'' being bit-wise XOR.

\begin{definition} For $1\le d\le n$, the $d$-\textsc{Hamming Distance} problem is to
compute the following Boolean function $\mathrm{HAM}_{n, d}: \{0,
1\}^n\times \{0, 1\}^n\to \{0, 1\}$, with $\mathrm{HAM}(x, y)=1$ if
and only if $|x\oplus y| > d$.
\end{definition}

\begin{lemma}[Yao]\label{yao}
$R^{\|, \mathrm{pub}}(\mathrm{HAM}_{n,d})=O(d^2)$.
\end{lemma}

In a recent paper \cite{GKW04}, Gavinsky, Kempe and de Wolf gave
another classical protocol, which is an improvement over Yao's
when $d\gg\log n$.

\begin{lemma}[GKW]\label{gavinsky} $R^{\|,pub}(\mathrm{HAM}_{n,d})=O(d\log n)$.
\end{lemma}

In this paper, we observe a lower bound for
$Q^*(\mathrm{HAM}_{n,d})$, which is also a lower bound for
$R^{||,pub}(\mathrm{HAM}_{n,d})$ according to Equality \eqref{eq:
relation}.

Notice that $\mathrm{HAM}(x, y)= n-\mathrm{HAM}(x, \bar y)$, where
$\bar y\defeq 11\cdots 1\oplus y$. Therefore
$Q^*(\mathrm{HAM}_{n,d})=Q^*(\mathrm{HAM}_{n,n-d})$, and we need
only consider the case $d\le n/2$.

\begin{proposition}\label{thm-lower} For any
$d\le n/2$, $Q^*(\mathrm{HAM}_{n,d})= \Omega(d)$.
\end{proposition}

We then construct a public-coin randomized SMP protocol that almost
matches the lower bound and improves both of the above protocols.

\begin{theorem}\label{thm-upper2}
$R^{||,pub}(\mathrm{HAM}_{n,d})=O(d\log d)$.
\end{theorem}

We shall prove the above two results in the following sections.
Finally we discuss open problems and a plausible approach for
closing the gap.

\vspace{1em} \noindent \textbf{Other related work: }Ambainis,
Gasarch, Srinavasan, and Utis \cite{AGSU04} considered the
\emph{error-free} communication complexity, and proved that any
\emph{error-free} quantum protocol for the Hamming Distance
problem requires at least $n-2$ qubits of communication in the
interactive model, for any $d\leq n-1$.

\section{Lower bound of the quantum communication complexity of the Hamming
Distance problem}\label{lowerbound} For proving the lower bound, we
restrict $\mathrm{HAM}_{n, d}$ on those pairs of inputs with equal
Hamming distance. More specifically, for an integer $k$, $1\le k\le
n$, define $X_k=Y_k\defeq \{x: x\in \booln, |x|=k\}$. Let
$\mathrm{HAM}_{n,k,d}: X_k\times Y_k\to \bool$ be the restriction of
$\mathrm{HAM}_{n, d}$ on $X_k\times Y_k$.

Before proving Proposition \ref{thm-lower}, we briefly introduce
some related results. Let $x, y\in \booln$. The
\textsc{Disjointness} problem is to compute the following Boolean
function $\mathrm{DISJ}_n: \{0, 1\}^n\times \{0, 1\}^n\to\{0,
1\}$, $\mathrm{DISJ}_n(x, y)=1$ if and only if there exists an
integer $i$, $1\le i\le n$, so that $x_i=y_i=1$. It is known that
$R(\mathrm{DISJ}_n)=\Theta(n)$ \cite{KS92} \cite{Razborov90}, and
$Q^*(\mathrm{DISJ}_n)=\Theta(\sqrt{n})$
\cite{Razborov03}\cite{AA03}.

We shall use an important lemma in Razborov\cite{Razborov03}, which
is more general than his remarkable lower bound on quantum
communication complexity of \textsc{Disjointness}. Here we may abuse
the notation by viewing $x\in\{0,1\}^n$ as the set $\{i\in [n]: x_i
= 1\}$.

\begin{lemma}[Razborov]\label{Razborov2}
Suppose $k\le n/4$ and $l\le k/4$. Let $D:[k]\to \bool$ be any
Boolean predicate such that $D(l)\neq D(l-1)$. Let
$f_{n,k,D}:X_k\times Y_k\to \bool$ be such that
$f_{n,k,D}(x,y)\defeq D(|x\cap y|)$. Then $Q^*(f_{n,k,D})=
\Omega(\sqrt{kl})$.
\end{lemma}

\beginproofof{Proposition \ref{thm-lower}}
Consider $D$ in Lemma \ref{Razborov2} such that $D(t)=1$ if and only
if $t< l$. For any $x,y\in X_k$, we have $|x\cap y|= k -
\mathrm{HAM}(x,y)/2$. Let $l=k-d/2$, then $k - \mathrm{HAM}(x,y)/2 <
l$ if and only if $\mathrm{HAM}(x,y)
> d$. Therefore, $D(|x\cap y|)=1$ if and only
if $\mathrm{HAM}(x,y)
> d$. This implies that $f_{n,k,D}$ and $\mathrm{HAM}_{n,k,d}$ are actually
the same function, and thus
$Q^*(f_{n,k,D})=Q^*(\mathrm{HAM}_{n,k,d})$.

To use lemma {\ref{Razborov2}}, the following two constraints on $k$
and $l$ need to be satisfied: $k\le n/4$ and $l\le k/4$. When $d\leq
3n/8$, let $k = 2d/3 \le n/4$, then $l = 2d/3 - d/2 = d/6 \le n/16$.
Both requirements for $k$ and $l$ are satisfied. So applying lemma
{\ref{Razborov2}}, we get $Q^*(\mathrm{HAM}_{n,k,d})=
Q^*(f_{n,k,D})=\Omega(\sqrt{kl})=\Omega(d)$.

For $3n/8< d\le n/2$, it is reduced to the above case ($d \leq
3n/8$) rather than lemma {\ref{Razborov2}}. Let $m = \lceil
8d/5-3n/5 \rceil$. Fix first $m$ bits in $x$ to be all $1$'s, and
use $x'$ to denote $x_{m+1}\dots x_n$. Similarly, fix first $m$ bits
of $y$ to be all $0$'s, and use $y'$ to denote $y_{m+1}\dots y_n$.
Put $n'=n-m$, $k'=n'/4$, and $d'=d-m$. Then
$\mathrm{HAM}(x,y)=\mathrm{HAM}(x',y')+m$ and
$Q^*(\mathrm{HAM}_{n,d})(x,y)\ge
Q^*(\mathrm{HAM}_{n',k',d'})(x',y')$. It is easy to verify that $d'
\leq 3n'/8$ and $d' = \Omega(d)$. Employing the result of the case
that $d \le 3n/8$, we have
$Q^*(\mathrm{HAM}_{n',k',d'})=\Omega(d')$. Thus
$Q^*(\mathrm{HAM}_{n,d})\ge
Q^*(\mathrm{HAM}_{n',k',d'})=\Omega(d')=\Omega(d)$. \finishproof

\section{Upper bound of the classical communication complexity of the Hamming
Distance problem} To prove theorem \ref{thm-upper2}, we reduce the
$\mathrm{HAM}_{n,d}$ problem to $\mathrm{HAM}_{16d^2, d}$ problem by
the following lemma.
\begin{lemma}\label{lemma-upper-reduction}
\[R^{||, \mathrm{pub}}(\mathrm{HAM}_{n,d}) = O(R^{||, \mathrm{pub}}(\mathrm{HAM}_{16d^2,d})) + O(d\log
d)\]
\end{lemma}

Note that Theorem \ref{thm-upper2} immediately follows from Lemma
$\ref{lemma-upper-reduction}$ because by Lemma $\ref{gavinsky}$,
$R^{||, \mathrm{pub}}(\mathrm{HAM}_{n,d})=O(d\log n)$, thus $R^{||,
\mathrm{pub}}(\mathrm{HAM}_{16d^2,d}) = O(d\log d^2) = O(d\log d)$.
Now by Lemma $\ref{lemma-upper-reduction}$, we have $R^{||,
\mathrm{pub}}(\mathrm{HAM}_{n,d})=O(d\log d)$. So in what follows,
we shall prove Lemma $\ref{lemma-upper-reduction}$. Define a partial
function $\mathrm{HAM}_{n,d|2d}(x,y)$ with domain $\{(x,y):\ x,y\in
\{0,1\}^n, \ |x\oplus y| \text{ is either less than $d$ or at least
$2d$} \}$ as follows.
\begin{equation}
\mathrm{HAM}_{n,d|2d}(x,y)= \left\{
\begin{array}{ll}
  0 & \textrm{If } \mathrm{HAM}(x,y)\le  d\\
  1 & \textrm{If } \mathrm{HAM}(x,y)> 2d \\
\end{array}
\right.
\end{equation}
Then
\begin{lemma}\label{lemma-upper1}
\[ R^{||,\mathrm{pub}}(\mathrm{HAM}_{n,d|2d})=O(1)\]
\end{lemma}
\beginproofof{Lemma \ref{lemma-upper1}}
We revise Yao's protocol \cite{Yao03} to design an $O(1)$ protocol
for $\mathrm{HAM}_{n,d|2d}$. Assume the Hamming distance between
$x$ and $y$ is $k$. Alice and Bob share some random public string,
which consists of a sequence of $\gamma n$($\gamma$ is some
constant to be determined later) random bits, each of which is
generated independently with probability $p=1/(2d)$ of being $1$.
Denote this string by $z_1, z_2, \cdots, z_{\gamma}$, each of
length $n$. Party $A$ sends the string $a=a_1a_2\cdots a_{\gamma
}$ to the referee, where $a_i= x\cdot z_i \pmod 2 $. Party $B$
sends the string $b=b_1b_2\cdots b_{\gamma}$ to the referee, where
$b_i= y\cdot z_i \pmod 2 $. The referee announces
$\mathrm{HAM}_{n,d}(x,y)=1$ if and only if the Hamming distance
between $a$ and $b$ is more than $m=(1/2-q)\gamma$ where
$q=((1-1/d)^d + (1-1/d)^{2d})/4$.

Now we prove the above protocol is correct with probability at
least $49/50$. Let $c_i=a_i \oplus b_i$. Notice that the Hamming
distance between $a$ and $b$ is the number of $1$'s in $c = c_1c_2
\cdots c_{\gamma}$. We need the following Lemma by Yao
\cite{Yao03}

\begin{lemma} Assume that the Hamming distance between $x$ and $y$ is $k$.
Given $c$ as defined above, each $c_i$ is an independent random
variable with probability $\alpha_k$ of being $1$, where $\alpha_k
= 1/2 - 1/2(1-1/d)^k$.
\end{lemma}

Since $\alpha_k$ is an increasing function over $k$, to separate
$k\le d$ from $k
> 2d$, it would be sufficient to discriminate the two cases that
$k=d$ and $k=2d$. Let $N_k$ be a random variable denoting the number
of $1$'s in $c$, and $E(N_k)$ and $\sigma(N_k)$ denote corresponding
expectation and standard deviation, respectively. Then we have
$E(N_k)= \alpha_k \gamma$, and $\sigma(N_k) \le (\alpha_k
\gamma)^{1/2}$. Thus $E(N_{2d})- E(N_d) =\gamma (\alpha_{2d} -
\alpha_{d}) ={1\over 2}\gamma(1-{1\over d})^d(1-(1-{1\over d})^d)
\ge {1\over 8}\gamma$. Let $\gamma=20000$, then $E(N_{2d})-E(N_d)\ge
2500$, while $\sigma(N_d), \sigma(N_{2d})< ({1\over
2}\gamma)^{1/2}=100$. The cutoff point in the protocol is the middle
of $E(N_d)$ and $E(N_{2d})$. By Chebyshev Inequility, with
probability of at most $1/100$, $|N_d -E(N_d)|> 10 \sigma(N_d)=
1000$. So does $N_{2d}$. Thus with probability of at least $49/50$,
the number of $1$'s in $c$ being more than cutoff point implies
$k>2d$ and vice versa. Therefore, $O(\gamma)$ communication is
sufficient to discriminate the case $\mathrm{HAM}(x,y)>2d$ and
$\mathrm{HAM}(x,y) \leq d$ with error probability of at most $1/50$.
\finishproof

The following fact is also useful
\begin{fact}\label{fact-upper}
If $2d$ balls are randomly thrown into $16d^2$ buckets, then with
probability of at least $7/8$, each bucket has at most one ball.
\end{fact}
\beginproofof{Fact $\ref{fact-upper}$}
There are ${2d \choose 2}$ pairs of balls. The probability of one
specific pair of balls falling into the same bucket is ${1\over
16d^2} \cdot {1\over 16d^2} \cdot 16d^2 = {1\over 16d^2}$. Thus
the probability of having a pair of balls in the same bucket is
upper bounded by ${1\over 16d^2} \cdot {2d \choose 2} <1/8$. Thus
Fact $\ref{fact-upper}$ holds. \finishproof

Now we are ready to prove Lemma $\ref{lemma-upper-reduction}$.

\beginproofof{Lemma $\ref{lemma-upper-reduction}$}
If $16d^2\ge n$, the $O(d\log n)$ communication protocol in Lemma
$\ref{gavinsky}$ would also be a $O(d\log d)$ protocol.

If $16d^2<n$, suppose we already have a protocol $P_1$ of $C$
communication to distinguish the cases $|x\oplus y| \leq d$ and $d <
|x\oplus y| \leq 2d$ with error probability at most $1/8$. Then we
can have a protocol of $C+O(1)$ communication for
$\mathrm{HAM}_{n,d}$ with error probability at most $1/4$. Actually,
by repeating the protocol for $\mathrm{HAM}_{n, d|2d}(x,y)$ several
times, we can have a protocol $P_2$ of $O(1)$ communication to
distinguish the cases $|x\oplus y| \leq d$ and $|x\oplus y| > 2d$
with error probability at most $1/8$. Now the whole protocol $P$ is
as follows. Alice sends the concatenation of $m_{A,1}$ and
$m_{A,2}$, which are her messages when she runs $P_1$ and $P_2$,
respectively. So does Bob send the concatenation of his two
corresponding messages $m_{B,1}$ and $m_{B,2}$. The referee then
runs protocol $P_i$ on $(m_{A,i}, m_{B,i})$ and gets the results
$r_i$. The referee now announces $|x\oplus y| \leq d$ if and only if
both $r_1$ and $r_2$ say $|x\oplus y| \leq d$.

It is easy to see that the protocol is correct. If $|x\oplus y| \leq
d$, then both protocols announces so with probability at least
$7/8$, and thus $P$ says so with probability at least $3/4$. If
$|x\oplus y| > d$, then one of the protocols gets the correct range
of $|x\oplus y|$ with probability at least $7/8$, and thus $P$
announces $|x\oplus y| > d$ with probability at least $7/8$ too.

Now it remains to design a protocol of $O(R^{||,
\mathrm{pub}}(\mathrm{HAM}_{16d^2,d}))$ communication to distinguish
$|x\oplus y| \leq d$ and $d < |x\oplus y| \leq 2d$. First we assume
that $n$ is divisible by $16d^2$, otherwise we pad some $0$'s to the
end of $x$ and $y$. Using the public random bits, Alice divides $x$
randomly into $16d^2$ parts evenly, Bob also divides $y$
correspondingly. Let $A_i, B_i$($1\le i\le 16d^2$) denote
corresponding parts of $x, y$. By Fact $\ref{fact-upper}$, with
probability at least $7/8$, each pair $A_i, B_i$ would contain at
most one bit on which $x$ and $y$ differ. Therefore, the Hamming
distance of $A_i$ and $B_i$ would be either $0$ or $1$, i.e, the
Hamming distance of $A_i$ and $B_i$ equals the parity of $A_i\oplus
B_i$, which is further equal to $\mathrm{PARITY}(A_i) \oplus
\mathrm{PARITY}(B_i)$. Let $a_i$ denote the parity of $A_i$, $b_i$
denote the parity bit of $B_i$, and let $a=a_1 a_2 \cdots
a_{16d^2}$, $b=b_1b_2 \cdots b_{16d^2}$. Then
$\mathrm{HAM}_{16d^2,d}(a,b)=\mathrm{HAM}_{n,d}(x,y)$ with
probability at least $7/8$. So we run the best protocol for
$\mathrm{Ham}_{16d^2,d}$ on the input $(a,b)$, and use the answer to
distinguish $|x\oplus y| \leq d$ and $d < |x\oplus y| \leq 2d$.
\finishproof

\section{Discussion}
We conjecture that our quantum lower bound in lemma
$\ref{thm-lower}$ is tight. It seems plausible to remove the $O(\log
d)$ factor in our upper bound. Recently, Aaronson and Ambainis
\cite{AA03} sharpened the upper bound of the Set Disjointness
problem from $O(\sqrt{n}\log n)$ to $O(\sqrt n)$ using quantum local
search instead of Grover's search. In their method, it takes only
constant communication qubits to synchronize two parties and
simulate each quantum query. From Yao's protocol \cite{Yao03}, one
can easily derive an $O(d\log d)$ two way interactive quantum
communication protocol using quantum counting\cite{BHT98} and the
connection between quantum query and communication \cite{BCW98}.
Methods similar to \cite{AA03} might help to remove the $O(\log d)$
factor in this upper bound.
\section{Acknowledgment}
We thank Alexei Kitaev for suggesting the relation of the
\textsc{Hamming Distance} problem and the \textsc{Disjointness}
problem, Ronald de Wolf for pointing out a mistake in our earlier
draft and anonymous reviewers for helping us improve the
presentation of this paper.
\bibliographystyle{plain}

\end{document}